\documentclass[%
reprint,
superscriptaddress,
 amsmath,amssymb,
 aps,
]{revtex4-2}

\usepackage{graphicx,setspace}
\usepackage{dcolumn}
\usepackage{bm}
\usepackage[mathlines]{lineno}
\usepackage{xcolor}
\usepackage[colorlinks=true,citecolor=teal,linkcolor=olive,anchorcolor=green,urlcolor=teal]{hyperref}
\usepackage{mathrsfs}
\usepackage{float}
\usepackage{ulem}
\usepackage[utf8]{inputenc}

\newcommand{\nn}{\nonumber}
\def\be{\begin{equation}}
\def\ee{\end{equation}}
\def\ba{\begin{eqnarray}}
\def\ea{\end{eqnarray}}

\def\nn{\nonumber}
\def\lf{\left}
\def\rt{\right}

\def\lf{\left}\def\rt{\right}   \def\y {\psi}   \def\p {\pi}        \def\k {\kappa}   \def\x {\xi}     \def\pd {\partial}\def\p {\pi}   
 \def\W{\Omega}             \def\grad{\nabla}\def\.{\cdot}
\def\math {\mathcal}
\begin{document}

\title{Strong cosmic censorship in near-extremal Kerr-Sen-de Sitter spacetime}
\author{Ming Zhang}
\email{mingzhang@jxnu.edu.cn}
\affiliation{Department of Physics, Jiangxi Normal University, Nanchang 330022, China}
\date{\today}
\author{Jie Jiang}
\email{Corresponding author. jiejiang@mail.bnu.edu.cn}
\affiliation{Department of Physics, Beijing Normal University, Beijing 100875, China}

\begin{abstract}
We first calculate equations of motion for particles in the Kerr-Sen-de Sitter black hole spacetime. Then in the eikonal regime, we analytically obtain the quasi-normal resonant modes of massless neutral scalar field perturbation and find its  imaginary part to be characterized by the surface gravity of the near-extremal Kerr-Sen-de Sitter black hole with Cauchy horizon approaching the event horizon. We further show that the Penrose strong cosmic censorship conjecture is thus respected in this spacetime with dilaton scalar field and axion pseudoscalar field.

\end{abstract}
\maketitle

\section{Introduction}
Classical General Relativity (GR) as a theory with deterministic nature can predict the future directed evolution of the spacetime. Penrose  strong cosmic censorship (SCC) conjecture \cite{Penrose:1969pc} asserts that the  Cauchy horizons (CH)  do not form, as beyond which generic asymptotically flat initial data should be future inextendible. Thus, SCC should be respected, or else GR loses. In the principle of the SCC, the Cauchy horizon of the black hole can not stably exist as the perturbations of the fields become singular there. As a result, the Cauchy horizon will become a singularity if SCC holds. In Christodoulou-Chruściel's modern version of SCC \cite{Christodoulou:2008nj}, it is stated that it should not be generally possible to extend the metric of the black hole spacetime continuously to cross the CH with Christoffel symbol which is locally square-integrable, even as the field equations' weak solution.  Due to the exponential blueshift effect in the interior of the black hole upon the signal sent by an exterior observer, the CH should be unstable. Therefore, asymptotically flat charged Reissner-Nordström or rotating Kerr black hole though with a Cauchy horizon respect the SCC \cite{Simpson:1973ua,Poisson:1990eh,Dafermos:2003wr}.

However, for a black hole immersed in the Universe with a positive cosmological constant, the faith of SCC becomes fuzzy \cite{Chambers:1994ap,Brady:1998au}. Usually, the strength of the perturbative field outside the black hole can be measured by the quasi-normal modes (QNMs) with complex frequencies whose imaginary part indicates the decay rate of the modes \cite{Kokkotas:1999bd,Berti:2009kk,Konoplya:2011qq}. As the massless scalar QNMs corresponding to the perturbations exterior to the event horizon exponentially decay sufficiently rapidly (known as the competing redshift effect, determined by the spectral gap corresponding to the imaginary part of the nonzero dominant QNM \cite{Hintz:2016gwb,Hintz:2016jak}) due to the existence of a cosmological horizon, being enough to counterbalance the blueshift effect (which is governed by the surface gravity of the black hole \cite{chandrasekhar1982crossing}), for the Reissner-Nordström-de Sitter (RNdS) black hole in the near extremal regime (CH being near to the event horizon),  the scalar field can be extended across the CH as the solution of the Klein-Gordon equation in the spacetime, thus the SCC was shown to be violated in Ref. \cite{Cardoso:2017soq}. Nevertheless, it was further clarified in Refs. \cite{Hod:2018dpx,Zhang:2019nye,Hollands:2019whz} that the claim proposed in Ref. \cite{Cardoso:2017soq} is erroneous. Particularly, it was pointed out that the study in Ref. \cite{Cardoso:2017soq} ignores the fact that charged black holes must be formed from the gravitational collapse of self-gravitating charged matter fields (not neutral matter fields) \cite{Hod:2018dpx}. Also, it was shown that the SCC can be restored if the non-linear evolution of charged scalar field is performed  \cite{Zhang:2019nye}.

Along the line, the  Kerr-de Sitter (KdS) black hole was shown to preserve the SCC against the scalar and gravitational perturbations \cite{Dias:2018ynt}.  In the charged KdS case, it was also proved that the SCC can be respected against massless neutral scalar fields perturbation \cite{Hod:2018lmi}. Though the SCC has been inspected extensively for the spherically black holes perturbed by scalar, fermion, electromagnetic, and gravitational field perturbations (see, for example, Refs.  \cite{Destounis:2020yav,Dias:2020ncd,Burikham:2020dfi,Mishra:2020gce,Rahman:2020guv,Liu:2019lon,Guo:2019tjy,Liu:2019rbq,Gan:2019ibg,Mishra:2019ged,Gim:2019rkl,Liu:2019lon,Dias:2018etb,Ge:2018vjq,Destounis:2018qnb,Mo:2018nnu,Cardoso:2018nvb,Hod:2019zoa,Hod:2020ktb}), the investigations of the rotating  cases seem to be limited \cite{Rahman:2018oso,Gwak:2019ttv,Miguel:2020uln,Dias:2019ery, Casals:2020uxa}. Besides, the SCC of spacetimes in Einstein gravity has been investigated comprehensively, there is much needing to be done in the modified gravities, being corrections to GR, following former trials \cite{Gan:2019jac,Destounis:2019omd}.

In this paper, we will investigate the SCC for the Kerr-Sen black hole \cite{Sen:1992ua} with a positive cosmological constant, i.e., the Kerr-Sen-de Sitter (KSdS) black hole  \cite{Chong:2004na,Birkandan:2015yda,Wu:2020cgf}. Comparing with the Kerr-Newman-de Sitter black hole, the KSdS black hole has distinct characteristics. It is algebraically type-A and owns additional dilaton scalar field and dual axion pseudoscalar field \cite{Burinskii:1995hk}. The physical motivation of our study is that the KSdS spacetime is a solution of low-energy effective field theory describing heterotic string theory. As discussed in Ref. \cite{Bernard:2016wqo}, the Universe may work as the string theory describes instead of the Einstein-Maxwell theory does.  The solution, involving an antisymmetric tensor  gauge field as well as a nontrivial dilaton field,  is qualitatively different from black holes in the ordinary Einstein-Maxwell gravity theory. We wonder that whether these fields affect the validity of the SCC. The organization of the paper is as follows.  As there are no existed equations of motion for particles in the KSdS spacetime in the literature, we will derive it in Sec. \ref{sec2}. In Sec. \ref{sec3}, we will first calculate the Lyapunov exponent of the null circular geodesics around the near-extremal KSdS black hole, obtaining its relation with the surface gravity of the black hole. Then in Sec. \ref{sec4} we will conduct the calculation of the condition of violating the SCC and use the relation between the QNMs and the null circular orbits in the eikonal limit to analytically show whether the SCC  can be respected by the near-extremal KSdS black hole perturbed by the scalar perturbation. Our conclusion and discussion will be given in the last section.

\section{Geodesics of particles around KS\lowercase{d}S black hole}\label{sec2} 
The effective action of the four-dimensional heterotic string field theory in the low-energy limit can be written as \cite{Sen:1992ua}
\ba\begin{aligned}
I_\text{bulk}=&\int_{\mathcal{M}}\star\left(R+\frac{4+e^{-\psi}+e^\psi (1+\chi^2)}{L^2}-\frac{1}{2}\grad^a\psi\grad_a\psi\right.\\&\left.-\frac{1}{2}e^{2\psi}\grad^a\chi\grad_a\chi-e^{-\psi}\math{F}\right)+\frac{\chi}{2}\int_\mathcal{M} \bm{F}\wedge\bm{F}\,,
\end{aligned}\ea
where a non-zero positive cosmological constant is included, as $\Lambda=3/L^2$, with $L$ the dS radius. $R$ is the Ricci scalar. $\psi$ is the dilaton scalar field, $\chi$ is the axion pseudoscalar field, which is related to a three-form antisymmetric tensor $H_{abc}$ by $\bm{H}=-e^{2 \psi} \star \rm{d} \chi$. $\bm{F}={\rm{d}}\bm{A}$ is the electromagnetic field tensor with $\bm{A}$ the gauge potential and we have denoted $\math{F}\equiv F_{ab}F^{ab}$. Varying the bulk action, the equations of motion read
\begin{equation}\label{eom1}
\begin{aligned}
&R_{ab}-\frac{1}{2}R g_{ab}-\frac{4+e^{-\psi}+e^\psi(1+\chi^2)}{2l^2}g_{ab}\\&=8\p\lf(T_{ab}^\text{A}+T_{ab}^\text{B}+T_{ab}^\text{DIL}\rt),
\end{aligned}
\end{equation}
\begin{equation}
\grad_a \tilde{F}^{ac}+\frac{1}{2}\lf(\tilde{H}^{abc}F_{ab}-A_a\grad_b \tilde{H}^{abc}\rt)=8\sqrt{2}\p j^c,
\end{equation}
\begin{equation}
E_\text{B}^{bc}=\frac{1}{32\p}\grad_a\tilde{H}^{abc}=0,
\end{equation}
\begin{equation}
E_\y=\frac{1}{16\p}\lf(\grad^2\y+\frac{1}{8}e^{-\y}F^2+\frac{1}{6}e^{-2\y}H^2\rt)=0,
\end{equation}
\begin{equation}
T_{ab}^\text{A}=\frac{e^{-\y}}{32\p}\lf(F_{ac}F_b{}^c-\frac{1}{4}g_{ab}F^2\rt),
\end{equation}
\begin{equation}
T_{ab}^\text{B}=\frac{e^{-2\y}}{32\p}\lf(H_{acd}H_b{}^{cd}-\frac{1}{6}g_{ab}H^2\rt),
\end{equation}
\begin{equation}
T_{ab}^\text{DIL}=\frac{1}{16\p}\lf(\grad_a\y\grad_b\y-\frac{1}{2}g_{ab}\grad_c\y\grad^c\y\rt).
\end{equation}
Here we have defined
\ba\begin{aligned}
\tilde{\bm{F}}=e^{-\y}\bm{F}\,,\ \ \tilde{\bm{H}}=e^{-2\y} \bm{H}\,.
\end{aligned}\ea

In the Boyer-Lindquist coordinates, the KSdS black hole solutions deriving from the above equations of motion read \cite{Chong:2004na,Birkandan:2015yda,Wu:2020cgf}
\begin{equation}
\begin{aligned}
ds^{2}=&-\frac{\Delta_{r}}{\Sigma} \left(\frac{dt}{I} - \dfrac{a}{I} \sin^2\theta d\phi \right)^2 +\Sigma  \left(\frac{dr^2}{\Delta_{r}}+\frac{d\theta^2}{\Delta_\theta} \right) \\&+ \frac{\Delta_\theta \sin^2\theta}{\Sigma} \left( \frac{a dt}{I} -\frac{(r^2+2br+a^2)}{I}d\phi \right)^2,\label{met10}
\end{aligned}
\end{equation}
\begin{equation}
\bm{A}=\frac{q r}{\Sigma}\left(d t-\frac{a \sin ^{2} \theta}{\Xi} d \varphi\right),
\end{equation}
\begin{equation}
\psi=\ln \left(\frac{r^{2}+a^{2} \cos ^{2} \theta}{\Sigma}\right),
\end{equation}
\begin{equation}
\chi=\frac{2 b a \cos \theta}{r^{2}+a^{2} \cos ^{2} \theta},
\end{equation}
where
\begin{eqnarray}
 \Delta_{r}&=&\left[1-\frac{\Lambda\left(r^2+2br\right)}{3}\right](r^2+2br+a^2)-2mr,\nn \\ \Delta_{\theta}&=& 1+\frac{\Lambda a^{2}}{3}\cos^{2}\theta,\nn \\
 I &=& 1+\frac{\Lambda a^{2}}{3},\nn \\ \Sigma &=& r^{2}+2br+a^{2}\cos^{2}\theta,\nn\\ b&=&q^2/(2 m)\nn.
\end{eqnarray}
$b$ is the twisted parameter, $M, q, a$ are individually the mass, $U(1)$ charge, and angular momentum per unit mass of the black hole.

The spacetime is pathological unless there are three positive roots for the blackening factor $\Delta_r$: the Cauchy horizon whose boundary is determined by the initial data, the event horizon and the cosmological horizon, with coordinate radii $r_-$, $r_+$, and $r_c$. Moreover, we have $r_{-} \leq r_{+} \leq r_{c}$. To guarantee the regularity of the horizons we must have $0 \leqslant a \leqslant a+b \leqslant m$. The black hole becomes extremal when $a+b=m$.  The Killing vector field which generates the inner and outer horizons is given by
\ba\begin{aligned}
\x_{\pm}^a=\left(\frac{\pd}{\pd t}\right)^a+\W_{\pm} \left(\frac{\pd}{\pd \y}\right)^a,
\end{aligned}\ea
in which 
$$\W_{\pm}=\left.-\frac{g_{t \phi} }{ g_{\phi \phi}}\right|_{r=r_\pm}=\frac{a}{r^2_{\pm}+2br_\pm+a^2}$$
relates to the angular velocities of the inner and outer horizons. Note that if there is not a rescaling factor $1/I$ for the term $dt$ in the metric (\ref{met10}), the angular velocities will be
$$
\W_{\pm}=\frac{a I}{r^2_{\pm}+2br_\pm+a^2}.
$$
The surface gravity, defined by $\x_{\pm}^b\grad_b\x_\pm^a=\k_{\pm}\x_{\pm}^b$, can be obtained as
\begin{equation}
\begin{aligned}
\kappa_\pm&=-\frac{1}{2} \lim _{r \rightarrow r_\pm} \sqrt{\frac{-g^{11}}{g^{00}}} \frac{\partial}{\partial r} \ln \left(-g^{00}\right)\\&=\left.\frac{\Delta '_r}{2 I \left[a^2+r (2 b+r)\right]}\right|_{r=r_\pm}.
\end{aligned}
\end{equation}

From the symmetries of the spacetime, which are characterized by the Killing vectors $\partial_t$ and $\partial_\phi$, we have the conserved energy $E$ and conserved angular momentum $L_z$,
\begin{align}
E &=-g_{t \mu} \dot{x}^{\mu}, \\
L_z &=g_{\phi \mu} \dot{x}^{\mu}.
\end{align}

The general form of the Hamilton-Jacobi equation for the particle, from which we can obtain the geodesics, reads \cite{Carter:1968rr}
\begin{equation}
\frac{\partial S}{\partial \lambda}=-\frac{1}{2} g^{\mu \nu} \frac{\partial S}{\partial x^{\mu}} \frac{\partial S}{\partial x^{\nu}},
\end{equation}
with $\lambda$ the affine parameter relating to the proper time $\tau$ by $\tau=\mu\lambda$ (the specific value of $\mu$ does not have significance so that we can set it to be unity) and $S$ the Jacobi action. Using the constants of motion, we may set the Jacobi action in a separable form as
\begin{equation}
S=-\frac{1}{2} \mu^{2} \lambda-E t+L_z \phi+S_{r}(r)+S_{\theta}(\theta).
\end{equation}
Then we have 
\begin{align}
d S_{r} / d r&=\Delta_r^{-1}\mathcal{V}_r(r), \\
d S_{\theta} / d \theta &=\sqrt{\Theta(\theta)},
\end{align}
where
\begin{align}
\mathcal{V}_r(r)&=\left[\left(r^{2}+2br+a^{2}\right) I E-a I L_z\right]^{2}-\Delta_r\left[\mu^{2} r^{2}+K\right], \\
\Theta(\theta)&=\mathcal{Q}-\cos ^{2} \theta\left(a^{2}\left(\mu^{2}-E^2\right)+L_z^{2} \sin ^{-2} \theta\right),\\
\mathcal{Q}&=K-I^{2}\left(a E-L_{z}\right)^{2}.
\end{align}
$\mathcal{V}_r(r)$, $\Theta(\theta)$ are individually the radial effective potential and the longitudinal effective potential. $\mathcal{Q}$ is the separation constant and $K$ is the fourth integral constant of geodesic motion besides the conserved energy, angular momentum and Hamiltonian $\mathcal{H}=-\mu^2/2$ \cite{Carter:1968rr}. 

The first-order differential form of the particle's motion is encoded in the equations
\begin{align}
\Sigma \dot{r}&=\sqrt{\mathcal{V}_r(r)}, \\
\Sigma \dot{\theta}&=\sqrt{\Theta (\theta)}, \\
\Sigma \dot{t}&=-a\left(a E \sin ^{2} \theta-L_z\right)-\left(r^{2}+2br+a^{2}\right) \Delta_r^{-1} P_r, \\
\Sigma \dot{\phi}&=-\left(a E-L_z \sin ^{-2} \theta\right)-a \Delta_r^{-1} P_r,
\end{align}
where the dot over a symbol means derivative relative to $\lambda$. For the massless photons on the  equatorial circular orbit , these equations reduces to
\begin{align}
\dot{r}&=\pm \mathcal{V}_{r}^{1 / 2}(r),\\
(r^{2}+2br) \dot{\phi}&=-I P_{\theta}+\frac{a I P_{r}}{\Delta_{r}},\\
(r^{2}+2br) \dot{t}&=-a I P_{\theta}+\frac{\left(r^{2}+2 b r+a^{2}\right) I P_{r}}{\Delta_{r}},
\end{align}
where
\begin{align}
\mathcal{V}_{r}(r)&=(r^2+2 b r)^{-2}\left[P_{r}^{2}-\Delta_{r}\left(m^{2} r^{2}+K\right)\right],\label{eq322}\\
P_{r}&=I E\left(r^{2}+2 b r+a^{2}\right)-a I L_{z},\\
P_{\theta}&=I\left(a E-L_{z}\right),\\
K&=I^{2}\left(a E-L_{z}\right)^{2}.
\end{align}

\section{Lyapunov exponent of  null circular geodesics around near-extremal KS\lowercase{d}S black hole}\label{sec3}

To characterize the instability timescale of the massless particle on the circular orbit, we use the Lyapunov exponent $\gamma$, which  is related to the effective potential and the coordinate time by \cite{Ferrari:1984zz,Cardoso:2008bp}
\begin{equation}
\gamma=\sqrt{\frac{\mathcal{V}_{r}^{\prime \prime}}{2 \dot{t}^{2}}},
\end{equation}
where the prime denotes the derivative with respect to the radial coordinate. We now use analytical techniques to calculate the Lyapunov exponent of the null circular orbit around the near-extremal KSdS black hole. It is not difficult to know that the circular orbit of the particle around the KSdS black hole locates on the equatorial plane \cite{Zhang:2020xub}. The radius $r_o$ of the null circular orbit is determined by the radial effective potential through the restrictions
\begin{align}
\mathcal{V}_{r}\left(r=r_{o}\right)&=0,\label{eq23}\\
\mathcal{V}_{r}^{\prime}\left(r=r_{o}\right)&=0.\label{eq24}
\end{align}
Substituting the expression of the effective potential Eq. (\ref{eq322}) into these equations, we obtain
\begin{equation}
\Delta_{r}\left(r=r_{o}\right)=\frac{\left[\left(r_{o}^{2}+2br_o+a^{2}\right) \Omega_{o}-a\right]^{2}}{\left(a \Omega_{o}-1\right)^{2}},\label{eq25}
\end{equation}
\begin{equation}
\Delta_{r}^{\prime}\left(r=r_{o}\right)=\frac{4( r_{o}+b) \Omega_{o} \cdot\left[\left(r_{o}^{2}+2br_o+a^{2}\right) \Omega_{o}-a\right]}{\left(a \Omega_{o}-1\right)^{2}},\label{eq26}
\end{equation}
where we have denoted $\Omega_{o}\equiv E/L_z$. Note that $\Omega_o$ is the angular velocity of the photon on the null circular orbit as $\Omega_o=\dot{\phi}/\dot{t}$.

In the near-extremal case where the Cauchy horizon approaches the event horizon, the location of the corotating circular orbit for the massless particle approaches the event horizon of the black hole \cite{Zhang:2018eau}, such that we have
\begin{align}
r_o-r_+&\ll 1,\\
\Delta_{r}^{\prime}\left(r=r_{o}\right)-\Delta_{r}^{\prime}\left(r=r_{+}\right)&\ll 1.\label{eq28}
\end{align}
The latter one further gives
\begin{equation}
\Omega_o-\Omega_+\ll 1.
\end{equation}
Based on these fact, we define two dimensionless parameters
\begin{align}
x &\equiv \frac{r_{o}-r_{+}}{r_{+}}\label{eq30},\\ 
y &\equiv \frac{\Omega_{o}-\Omega_+}{\Omega_+}.\label{eq31}
\end{align}
Using them, we have the near-horizon expansions of $\Delta_{r}\left(r=r_{o}\right)$ and $\Delta_{r}^{\prime}\left(r=r_{o}\right)$ as
\begin{equation}
\begin{aligned}
\Delta_{r}\left(r=r_{\mathrm{c}}\right)=&r_{+} \Delta_{r}^{\prime}\left(r=r_{+}\right) \cdot x\\&+\frac{1}{2} r_{+}^{2} \Delta_{r}^{\prime \prime}\left(r=r_{+}\right) \cdot x^{2}+\mathcal{O}\left(x^{2}\right),\label{eq32}
\end{aligned}
\end{equation}
\begin{equation}
\begin{aligned}
\Delta_{r}^{\prime}\left(r=r_{\mathrm{c}}\right)=&\Delta_{r}^{\prime}\left(r=r_{+}\right)\\&+r_{+} \Delta_{r}^{\prime \prime}\left(r=r_{+}\right) \cdot x+\mathcal{O}\left(x\right).\label{eq33}
\end{aligned}
\end{equation}
According to Eqs. (\ref{eq30}), (\ref{eq31}), (\ref{eq32}) and (\ref{eq33}), we can rewrite Eqs. (\ref{eq25}) and (\ref{eq26}) as
\begin{equation}
\begin{aligned}
&\frac{a^2 \left[a^2 y+2 b r_+x+2r_+^2  x)\right]^2}{\left[a^2+r_+ (2 b+r_+)\right]^2}\\&=\left[r_{+} \Delta_{r}^{\prime}\left(r=r_{+}\right) \cdot x+\frac{1}{2} r_{+}^{2} \Delta_{r}^{\prime \prime}\left(r=r_{+}\right) \cdot x^{2}\right] \\&\quad\cdot\left(a \Omega_{o}-1\right)^{2}[1+\mathcal{O}(x, y)],
\end{aligned}
\end{equation}
\begin{equation}
\begin{aligned}
&\frac{a\left[a^2 y+2 b r_+x+2r_+^2  x)\right]}{\left[a^2+r_+ (2 b+r_+)\right]}\\&=\left[ \Delta_{r}^{\prime}\left(r_{+}\right) + r_{+} \Delta_{r}^{\prime \prime}\left(r_{+}\right) \cdot x\right] \cdot\frac{\left(a \Omega_{o}-1\right)^{2}}{4 r_o \Omega_o} \cdot[1+\mathcal{O}(x, y)].
\end{aligned}
\end{equation}
Then we can get the expressions of the parameters $x$ and $y$ in terms of black hole parameters as
\begin{equation}
x=\left.\frac{2 \sqrt{2} a \Delta '_r}{r\Delta ''_r \sqrt{ 8 a^2-(2 b+r)^2 \Delta ''_r}}-\frac{\Delta '_r}{r \Delta ''_r}\right|_{r=r_+},\label{eq39}
\end{equation}
\begin{equation}
\begin{aligned}
y=&\frac{r (2 b+r)^2 \Delta '_r}{\sqrt{2} a^3 \sqrt{8 a^2-(2 b+r)^2 \Delta ''_r}}+\frac{2 (b+r) \Delta '_r}{a^2 \Delta ''_r}\\&\left.-\frac{4 \sqrt{2} (b+r) \Delta '_r}{a \Delta ''_r \sqrt{8 a^2-(2 b+r)^2 \Delta ''_r}}\right|_{r=r_+}.
\end{aligned}
\end{equation}
With the specific expressions of $x$ and $y$, we now calculate the Lyapunov exponent for the near-horizon null circular orbit. The second-order derivatives of the effective potential can be specified as
\begin{equation}\begin{aligned}\label{effd2}
&\mathcal{V}_{r}^{\prime \prime}(r=r_o)\\&=\frac{I^{2} L_{z}^{2}\left\{\left[\Omega_{o}\left(r_o^{2}+2 b r_o+a^{2}\right)-a\right]^{2}-\Delta_{ro}\left(a \Omega_{o}-1\right)^{2}\right\}^{\prime \prime}}{(r_o^2+2br_o)^2}\\&=\frac{I^{2} L_{z}^{2}}{(r_o^2+2br_o)^2}\left\{4 \Omega_o  \left(a^2 \Omega_o -a+\Omega_o  \left(2 b^2+6 b r_o+3 r_o^2\right)\right)\right.\\& \quad\left.-(a \Omega -1)^2 \Delta_{ro}^{\prime\prime}\right\}\\&=\frac{I^2 L_z^2 \left[8 a^2 \left(b+r_+\right){}^2-r_+^2 \left(2 b+r_+\right){}^2 \Delta ''_r\left(r_+\right)\right]}{r_+^2 (2 b+r_+)^2 \left(a^2+2 b r_++r_+^2\right)^2}\\&\quad\cdot[1+\mathcal{O}(x, y)],
\end{aligned}\end{equation}
where we have denoted $\Delta_{ro}=\Delta_{r}(r=r_o)$ and used the relation
\begin{equation}
\begin{aligned}
&\left[\Omega_{o}\left(r_o^{2}+2 b r_o+a^{2}\right)-a\right]^{2}-\Delta_{ro}\left(a \Omega_{o}-1\right)^{2}\\&=\left\{\left[\Omega_{o}\left(r_o^{2}+2 b r_o+a^{2}\right)-a\right]^{2}-\Delta_{ro}\left(a \Omega_{o}-1\right)^{2}\right\}^{\prime}=0,
\end{aligned}
\end{equation}
yielding from Eqs. (\ref{eq23}) and (\ref{eq24}). 

For the derivative of the coordinate time with respect to the affine parameter on the null circular orbit, we have 
\begin{equation}
\begin{aligned}
&\dot{t}^{-1}(r=r_{o})\\&=\frac{r^{2}+2br}{I^{2} L_{z}}\left\{-a\left(a \Omega_{o}-1\right)\right.\\&\quad\left.+\left.\frac{r^{2}+2 b r+a^{2}}{\Delta_{r}}\left[\Omega_{o}\left(r^{2}+2 b r+a^{2}\right)-a\right]\right\}^{-1}\right|_{r=r_o}\\&=\frac{r_{o}^{2}+2br_o}{I^{2} L_{z}\left(1-a \Omega_{o}\right)}\left[a+\frac{r_{o}^{2}+2 b r_o+a^{2}}{\sqrt{\Delta_{r}\left(r=r_{o}\right)}}\right]^{-1}\\&=\frac{(r_{+}+2b) \Delta_{r}^{\prime}\left(r_{+}\right)}{I^{2} L_{z} \sqrt{16 a^{2}-2 (r_{+}+2 b)^{2} \Delta_{r}^{\prime \prime}(r_+)}} \cdot[1+\mathcal{O}(x, y)],
\end{aligned}
\end{equation}
where in the second step we have used the relation (\ref{eq25}), in the third step we have used Eq. (\ref{eq32}) , and in the last step the Eqs. (\ref{eq31}) and (\ref{eq39}).

At last, the Lyapunov exponent of the null circular orbit around the near-extremal KSdS black hole can be obtained as
\begin{equation}\label{eq44}
\begin{aligned}
\gamma&=\frac{\Delta '_r (r_+)}{2 I r_+ \left[a^2+r_+ (2 b+r_+)\right]}\\&\quad\cdot \sqrt{\frac{r_+^2 (2 b+r_+)^2 \Delta ''_r-8 a^2 (b+r_+)^2}{(2 b+r_+)^2 \Delta ''_r-8 a^2}}\cdot[1+\mathcal{O}(x, y)]\\&=\frac{1}{r_+} \sqrt{\frac{r_+^2 (2 b+r_+)^2 \Delta ''_r-8 a^2 (b+r_+)^2}{(2 b+r_+)^2 \Delta ''_r-8 a^2}}\\&\quad\cdot\kappa_+ [1+\mathcal{O}(x, y)],
\end{aligned}
\end{equation}
where the equality is fulfilled when the twisted parameter $b$ vanishes.

\section{Strong cosmic censorship in near-extremal Kerr-Sen-de Sitter spacetime}\label{sec4}
The validity of SCC is closely related to the late time behavior of the linear field perturbation to the black hole, characterized by the QNMs. Here we study the massless scalar field as a toy example for the gravitational perturbations \cite{Cardoso:2017soq}. The QNMs of the massless scalar  are governed by the Klein-Gordon equation
\begin{equation}
\square \Phi=0,\label{eq45}
\end{equation}
together with specific causality boundary conditions which single out the quantized discret spectrum of quasi-normal frequencies. After performing the ansatz
\begin{equation}
\Phi_{n l m}(t, r, \theta, \phi)=e^{-i \omega t}  R_{n l m}(r) \Theta_{n l m}(\theta) e^{-i m \phi},
\end{equation}
with $\omega$  being the quasi-normal frequency, integers $n\,,l\,,m$ ($n$ is the multipole number, or overtone number, $l$ the spheroidal harmonic index, or the angular momentum of the scalar perturbation, and $m$ the azimuthal
harmonic index) labeling each mode. In the rotating KSdS case, the radial ordinary differential equation extracting from Eq. (\ref{eq45}) is
\begin{equation}
\left(\frac{\mathrm{d}^{2}}{\mathrm{~d} r_{*}^{2}}+\left(\omega-m \Omega_{\mathrm{BH}}(r)\right)^{2}-V_{\mathrm{BH}}(r)\right) R(r)=0,
\end{equation}
where we have defined the tortoise coordinate by
\begin{equation}
\mathrm{d} r_{*}=\frac{\Sigma\left(r^{2}+2b r+a^{2}\right)}{\Delta_{r}} \mathrm{d} r,
\end{equation}
and the potential function satisfies $V_{\mathrm{BH}}\left(r_c\right)=V_{\mathrm{BH}}\left(r_{+}\right)=V_{\mathrm{BH}}\left(r_{-}\right)=0$ and $\Omega_{\mathrm{BH}}\equiv a/(r^2+2 b r+a^2)$. 

The quasi-normal resonant modes should be purely ingoing at the event horizon of the KSdS black hole and purely outgoing at the cosmological horizon, that is, 
\begin{align}
R\left(r_{*}\to -\infty\right) &\sim e^{-i\left(\omega-m \Omega_{\mathrm{BH}}\left(r_{+}\right)\right) r_{*}}, \\
R\left(r_{*}\to\infty\right) &\sim e^{i\left(\omega-m \Omega_{\mathrm{BH}}\left(r_{c}\right)\right) r_{*}}.
\end{align}

As discovered in Ref. \cite{Cardoso:2017soq}, there will be three families of QNMs, namely the photon sphere (PS) modes related to the null circular geodesics, the purely imaginary dS modes described by the surface gravity at the cosmological horizon of the purely de Sitter spacetime, and the near-extremal modes dominating the dynamics in the limit that the event horizon  and Cauchy horizon approach each other. We here will consider the PS modes of the massless scalar field coupled with the near extremal KSdS black hole in the limit that the Cauchy horizon approaches to the event horizon. In the eikonal limit with $l= |m| \gg 1$, using the WKB mehtod \cite{Iyer:1986np,Konoplya:2019hlu}, the quasi-normal frequencis of the PS modes can be found to relate with the null circular geodesics through the relation \cite{Cardoso:2008bp,Konoplya:2017wot}
\begin{equation}\label{eq50}
\omega_{\mathrm{WKB}} \approx m \Omega_{o}^{+}-i\left(n+\frac{1}{2}\right) \gamma^{+},
\end{equation}
where 
\begin{equation}
\Omega_{o}^{+} \equiv \left.\frac{\dot{\phi}}{\dot{t}}\right|_{r=r_{+}}=\frac{a}{r_{+}^{2}+2br_++a^{2}}
\end{equation}
is the angular velocity of the photon on the corotating null circular orbit with a Lyapunov exponent $\gamma^+$ evaluated by Eq. (\ref{eq44}). We get the fundamental mode when the overtone number $n=0$. 

The fate of the SCC depends on the relation between the spectral gap $\alpha$ and the surface gravity $\kappa_-$ at the Cauchy horizon, with which we can define a control parameter 
\begin{equation}\label{eq52}
\beta\equiv\alpha/\kappa_-, 
\end{equation}
with $\alpha \equiv-\operatorname{Im}(\omega_{\mathrm{WKB}})$ \cite{Maeda:1999sv,Dafermos:2012np}. In fact, we have  \cite{bony2008decay,dyatlov2012asymptotic}
\begin{equation}
\left|\Phi-\Phi_{0}\right| \leq C e^{-\alpha t},
\end{equation}
with $\Phi_{0} \in \mathbb{C}$ being a constant shift, where $\Phi$ is a linear scalar perturbation. Thus the spectral gap is the size of the QNM-free strip below the real axis \cite{Cardoso:2017soq}.
 In the Einstein gravity, it has been suggested that SCC is violated if $\beta>1/2$ \cite{Costa:2014yha,Hintz:2015jkj,Cardoso:2017soq}. But in the modified gravity, we should to reconsider the criteria \cite{Destounis:2019omd}. According to the field equation (\ref{eom1}), to judge the extendibility of the black hole solution beyond the Cauchy horizon, we need 
\begin{equation}\begin{aligned}
&\int_{\mathcal{V}} d^{4} x \sqrt{-g}\left[G_{ab}-\frac{4+e^{-\psi}+e^\psi(1+\chi^2)}{2l^2}g_{ab}\right]\bar{\psi}\\&-8\pi\int_{\mathcal{V}} d^{4} x \sqrt{-g} \lf(T_{ab}^\text{A}+T_{ab}^\text{B}+T_{ab}^\text{DIL}\rt)\bar{\psi}=0
\end{aligned}\end{equation}
to be finite, where $\bar{\psi}$ is some test function, $\mathcal{V} \subset \mathcal{M}$ with $\mathcal{M}$ the spacetime manifold, and $G_{ab}\equiv R_{ab}-R g_{ab}/2$. As 
\begin{equation}\begin{aligned}
&\int_{\mathcal{V}} d^{4} x \sqrt{-g}\left[G_{ab}-\frac{4+e^{-\psi}+e^\psi(1+\chi^2)}{2L^2}g_{ab}\right] \bar{\psi}\\&\sim\int_{\mathcal{V}} d^{4} x \sqrt{-g}\left[\partial \Gamma+\Gamma^{2}-\frac{4+e^{-\psi}+e^\psi(1+\chi^2)}{2L^2}g_{ab}\right]\bar{\psi}\\&\sim -\int_{\mathcal{V}} d^{4} x \sqrt{-g}(\partial \bar{\psi}) \Gamma+\int_{\mathcal{V}} d^{4} x \sqrt{-g} \Gamma^{2}\bar{\psi}\\&\quad-\int_{\mathcal{V}}d^4 x \frac{4+e^{-\psi}+e^\psi(1+\chi^2)}{2L^2}\sqrt{-g}g_{ab}\bar{\psi},
\end{aligned}\end{equation}
where $\Gamma$ is the Christoffel symbol and the expansion $G_{\mu \nu} \sim \Gamma^{2}+\partial \Gamma$ are used, we need $\Gamma \in L_{\mathrm{loc}}^{2}$ with $L_{\mathrm{loc}}^{2}$ the space consisting of square integrable functions locally in $\mathcal{V}$ \cite{Costa:2017tjc,Dafermos:2017dbw,Destounis:2019omd}.
For the part of the energy-momentum tensor, we have 
\begin{equation}
\begin{aligned}
&\int_{\mathcal{V}} d^{4} x \sqrt{-g}\left( T_{ab}^\text{B}+T_{ab}^\text{DIL}\right) \bar{\psi} \\&\sim \int_{\mathcal{V}} d^{4} x \sqrt{-g}\left[(\partial \psi)^{2}+(\partial \chi)^{2}\right] \bar{\psi}.
\end{aligned}
\end{equation}
This means that to make the $(\partial \psi)^{2}$ and $(\partial \chi)^{2}$ integrable,  we must have $\psi \in H_{\mathrm{loc}}^{1}$ and $\chi \in H_{\mathrm{loc}}^{1}$, with $H_{\mathrm{loc}}^{p}$ the Sobolev space of functions in $L_{\mathrm{loc}}^{2}$ and $p$ means the order of the derivatives \cite{Dafermos:2017dbw}.
As the electromagnetic potential is regular at the Cauchy horizon, the integral of the energy-momentum for the electromagnetic field is finitely bounded.

As the integrable requirement is up to $H_{\mathrm{loc}}^{1}$, we know that for the KSdS spacetime, the control parameter determining the fate of the SCC is the same to the one in the Einstein case. In the eikonal limit, according to Eqs. (\ref{eq44}), (\ref{eq50}) and (\ref{eq52}), we have
\begin{equation}
\begin{aligned}
\beta&\leqslant \frac{-\operatorname{Im}(\omega_{\mathrm{WKB}})}{\kappa_+}\\&=\frac{1}{2r_+} \sqrt{\frac{r_+^2 (2 b+r_+)^2 \Delta ''_r-8 a^2 (b+r_+)^2}{(2 b+r_+)^2 \Delta ''_r-8 a^2}},
\end{aligned}
\end{equation}
where we have used the relation $\kappa_+\leqslant\kappa_-$ \cite{chandrasekhar1982crossing,Brady:1998au,SantiagoGerman:2003hp} and the information of the dilaton scalar field $\psi$ and axion pseudoscalar field $\chi$ are encoded into the expression. According to Eqs. (\ref{eq39}) and (\ref{effd2}), we know
\begin{align}
&&r_+^2 (2 b+r_+)^2 \Delta ''_r-8 a^2 (b+r_+)^2<0,\\&&(2 b+r_+)^2 \Delta ''_r-8 a^2<0.
\end{align}
 Besides, we have
\begin{equation}
\begin{aligned}
&r_+^2 (2 b+r_+)^2 \Delta ''_r-8 a^2 (b+r_+)^2\\&-r_+^2\left[(2 b+r_+)^2 \Delta ''_r-8 a^2\right]=-8 a^{2} b(b+2 r_+)\leqslant 0,
\end{aligned}
\end{equation}
with the equality fulfilled for $b=0$. So we get the minimal value of the polynomial
\begin{equation}
\left(\frac{1}{2r_+} \sqrt{\frac{r_+^2 (2 b+r_+)^2 \Delta ''_r-8 a^2 (b+r_+)^2}{(2 b+r_+)^2 \Delta ''_r-8 a^2}}\right)_{\rm{min}}=\frac{1}{2},
\end{equation}
which gives $\beta\leqslant 1/2$.
As a result, the SCC is shown to be respected by the near-extremal KSdS black hole with dilaton scalar field and axion pseudoscalar field. Note that $(2 b+r_+)^2 \Delta ''_r-8 a^2<0$ demands
\begin{equation}
\bar{a}>\frac{1}{2}+\bar{b}-\frac{1}{48} (2 \bar{b}+1)(2 \bar{b}^2+6 \bar{b}+3)\bar{\Lambda}+\mathcal{O}(\bar{\Lambda}^2),
\end{equation}
where $\bar{a}\equiv a/r_+,\,\bar{b}\equiv b/r_+,\,\bar{\Lambda}\equiv\Lambda r_+^2$ are dimensionless parameters.

\section{Conclusion and Discussion}\label{sec4}
Under perturbations of massless neutral scalar fields, we investigated the SCC in the near-extremal KSdS spacetime whose Cauchy horizon approaches the event horizon. To this end, we first calculated  the equations of motion for the photons and then analytically presented the Lyapunov exponent of the null circular orbit, expressing it in terms of the surface gravity of the near-extremal black hole. In the eikonal regime, after analyzing the critical control parameter above which the SCC will be violated, we further  proved that, even with the dilaton scalar field $\psi$ and axion pseudoscalar field $\chi$, the SCC is respected by the near-extremal KSdS black hole coupled with scalar fields, as the integrable requirement is up to $H_{\mathrm{loc}}^{1}$ and not affected by the fields.

The present result obtained in the paper is based on  analytical calculations. It is interesting to check it with numerical methods \cite{Jansen:2017oag}. We should also mention  efforts of consilidating the validity of SCC, both at the classical level \cite{Dafermos:2018tha} and at the quantum level \cite{Hollands:2019whz,Hollands:2020qpe}. For the former  case, SCC can be recovered if the initial data is allowed to be non-smooth; for the latter one, it was shown that the quantum stress tensor at the Cauchy horizon is sufficiently irregular. It is meaningful to extend the temporary work to consider the case with generic initial data on the CH as well as the quantum instability of the CH in the KSdS spacetime.

\section*{Acknowledgements}
M. Z. is supported by the National Natural Science Foundation of China (Grant No. 12005080).  J. J.  is supported by the National Natural Science Foundation of China (Grants No. 11775022 and No. 11873044).

\end{document}